\documentclass[aip,rsi,amsmath,amssymb,reprint]{revtex4-2}

\usepackage{graphicx}
\usepackage{dcolumn}
\usepackage{bm}

\usepackage[utf8]{inputenc}
\usepackage[T1]{fontenc}
\usepackage{mathptmx}
\usepackage{etoolbox}
\usepackage{siunitx}
\DeclareSIUnit{\gauss}{G}
\usepackage{subcaption}
\usepackage{float}
\usepackage{enumitem}
\usepackage{physics}
\usepackage{xcolor}
\usepackage{hyperref}
\usepackage[capitalise]{cleveref}

\newcommand{\Shalf}{\ensuremath{5s\,^{2}S_{1/2}}}
\newcommand{\Phalf}{\ensuremath{5p\,^{2}P_{1/2}}}

\newcommand{\Dthreehalf}{\ensuremath{4d\,^{2}D_{3/2}}}
\newcommand{\Dfivehalf}{\ensuremath{4d\,^{2}D_{5/2}}}
\newcommand{\SrIon}{\ensuremath{{}^{88}\mathrm{Sr}^{+}}}
\newcommand{\FWHMG}{\ensuremath{\Gamma_{G}}}
\newcommand{\YbIon}{\ensuremath{{}^{171}\mathrm{Yb}^{+}}}

\makeatletter
\def\@email#1#2{%
 \endgroup
 \patchcmd{\titleblock@produce}
  {\frontmatter@RRAPformat}
  {\frontmatter@RRAPformat{\produce@RRAP{*#1\href{mailto:#2}{#2}}}\frontmatter@RRAPformat}
  {}{}
}%
\makeatother
\begin{document}

\title[A trap-based platform for cooling and laser spectroscopy of fast ions toward radioactive-ion applications]{A trap-based platform for cooling and laser spectroscopy of fast ions toward radioactive-ion applications}
\author{Stefanos \surname{Pelonis}}
\affiliation{Instituut voor Kern- en Stralingsfysica, KU Leuven, Leuven, Belgium}
\email{stefanos.pelonis@kuleuven.be}

\author{Tobias \surname{Christen}}
\affiliation{Instituut voor Kern- en Stralingsfysica, KU Leuven, Leuven, Belgium}

\author{Wouter \surname{Gins}}
\affiliation{Instituut voor Kern- en Stralingsfysica, KU Leuven, Leuven, Belgium}

\author{Bram \surname{Hendrickx}}
\affiliation{Instituut voor Kern- en Stralingsfysica, KU Leuven, Leuven, Belgium}

\author{Phillip \surname{Imgram}}
\altaffiliation[Current address: ]{Facility for Antiproton and Ion Research in Europe (FAIR) GmbH, Planckstr. 1, 64291 Darmstadt, Germany}
\affiliation{Instituut voor Kern- en Stralingsfysica, KU Leuven, Leuven, Belgium}

\author{Angelos \surname{Karadimas}}
\affiliation{Instituut voor Kern- en Stralingsfysica, KU Leuven, Leuven, Belgium}

\author{\'Agota \surname{Koszor\'us}}
\affiliation{Instituut voor Kern- en Stralingsfysica, KU Leuven, Leuven, Belgium}
\affiliation{Belgian Nuclear Research Centre, SCK CEN, Mol, Belgium}

\author{Sandro \surname{Kramer}}
\affiliation{Instituut voor Kern- en Stralingsfysica, KU Leuven, Leuven, Belgium}

\author{Stef \surname{Kunnen}}
\affiliation{Instituut voor Kern- en Stralingsfysica, KU Leuven, Leuven, Belgium}

\author{Pierre \surname{Lassegues}}
\altaffiliation[Current address: ]{Laboratoire National des Champs Magn\'etiques Intenses, 143 Av. de Rangueil, 31400 Toulouse, France}
\affiliation{Instituut voor Kern- en Stralingsfysica, KU Leuven, Leuven, Belgium}

\author{Andrea \surname{Raggio}}
\affiliation{Instituut voor Kern- en Stralingsfysica, KU Leuven, Leuven, Belgium}

\author{Robbe \surname{Van Duyse}}
\affiliation{Instituut voor Kern- en Stralingsfysica, KU Leuven, Leuven, Belgium}

\author{Ruben P. \surname{de Groote}}
\affiliation{Instituut voor Kern- en Stralingsfysica, KU Leuven, Leuven, Belgium}

\date{\today}

\begin{abstract}
We report on the first experiments with the Stopping and Trapping of Radioactive Isotopes for Precision Experiments (STRIPE) segmented linear Paul trap. The intent for this apparatus is to serve as an efficient capture and laser-cooling trap for radioactive isotopes produced at radioactive ion beam facilities. We demonstrate the deceleration of \SI{10}{\kilo\electronvolt} ion beams and their trapping and laser cooling into Coulomb crystals. As an initial spectroscopic benchmark, the dipole-forbidden $\Shalf \rightarrow \Dfivehalf$ transition in \SrIon{} was driven, partially resolving the Zeeman structure and yielding a Gaussian linewidth $\FWHMG = \SI{14.5(8)}{\mega\hertz}$, with the residual width attributed to the ions' micromotion. The lifetime of the metastable \Dfivehalf{} state was measured to be $\tau_{\mathrm{obs}} = 0.325(5)_{\mathrm{stat}}(4)_{\mathrm{sys}}\,\si{\second}$, uncorrected for collisional quenching. Several measurements to quantify the cooling timescale, the detection rate of the photon collection region and the trap storage lifetime were also performed. The developed technique will allow radioactive ion beams to be decelerated and cooled for precision spectroscopy measurements.
\end{abstract}

\maketitle

\section{\label{sec:Introduction}INTRODUCTION}

Laser spectroscopy (LS) gives nuclear-model-independent access to four complementary nuclear observables: the spin $I$, the magnetic dipole and electric quadrupole moments $\mu$ and $Q_s$~\cite{Neyens2003}, and the mean-square charge radius $\langle r^2 \rangle$~\cite{Yang2023}. For radioactive isotopes it is typically performed on fast beams in the collinear geometry~\cite{Cheal_2010,Neugart2017,Koszorus2024}, where the microsecond-scale ion--laser interaction time restricts it mostly to dipole-allowed transitions with natural linewidths of tens of \si{\mega\hertz}. Confining the ions in a Paul trap extends the interaction time by many orders of magnitude and makes narrow dipole-forbidden and RF transitions accessible~\cite{Blaum_2013}. These enable both a higher precision on the above observables and access to higher-order ones, such as the magnetic octupole moment, an additional probe of the distribution of nuclear magnetization~\cite{deGroote2022}, and isotope shifts precise enough for King-plot searches for physics beyond the Standard Model~\cite{Berengut2025}. Applying this to radioactive isotopes requires injecting externally produced ions, delivered as a beam from a radioactive ion beam (RIB) facility, into a trap and laser-cooling them~\cite{deGroote2024}.

To date, transition frequencies of stable, laser-cooled ions in Paul traps have been measured with sub-hertz uncertainty by several groups, for example in \SrIon{} and \YbIon{}~\cite{Ludlow2015}. Isotope shifts have reached hertz-level precision along the even Yb$^{+}$ chain~\cite{Door2025}, and uncertainties below \SI{100}{\milli\hertz} along the even Ca$^{+}$ chain~\cite{Wilzewski2025} and between $^{86,88}$Sr$^{+}$~\cite{Manovitz2019}. However, the application of such measurements to radioactive cases has been limited. Typical examples include the work on $^{224}\mathrm{Ra}^{+}$ and $^{133}\mathrm{Ba}^{+}$~\cite{Fan2023,PhysRevLett.119.100501}, where the radioactive samples were placed close to the linear Paul trap and ion loading was performed by photoionization. Ra$^+$ has also been decelerated and captured from \si{\kilo\electronvolt} energies using buffer gas~\cite{NunezPortela2014}, where collisions leading to thermalization at room temperature limited the precision to the \si{\giga\hertz} level. A similar approach was used for $^{11}\mathrm{Be}^{+}$~\cite{Takamine11Be}, where the buffer gas was removed from the trap prior to measurement. This improved the precision to the \si{\kilo\hertz} level, at the expense of a delay of several seconds before each measurement. Avoiding buffer gas altogether would extend in-trap LS to reactive species, for which molecule formation can prevent spectroscopy, and to highly charged ions, which would otherwise rapidly undergo charge exchange. Within this context, capture and cooling of \si{\kilo\electronvolt} beams without buffer gas has been demonstrated by Sels \textit{et al.}~\cite{PhysRevResearch.4.033229} for Mg$^+$, K$^+$ and O$_2^+$, using Doppler and sympathetic cooling.

Following the same approach, we recently showed from time-of-flight (ToF) spectra that the present trap can decelerate, capture and laser-cool \SI{10}{\kilo\electronvolt} $^{88}$Sr$^+$ ions without buffer gas~\cite{Imgram2025}. The present work adds optical detection of the trapped ions, with a charge-coupled device (CCD) camera and a photomultiplier tube (PMT). The camera shows that the ions form resolved Coulomb crystals, placing their temperature close to the Doppler limit, and the PMT gives the photon detection rate per ion. The trap storage lifetime has also been improved by more than an order of magnitude over Ref.~\cite{Imgram2025}. Together, these enabled driving the $\Shalf \rightarrow \Dfivehalf$ electric-quadrupole (E2) transition at \SI{674}{\nano\metre} as a first spectroscopic benchmark, from which we gauge the micromotion of the ions in the trap.

This paper is structured as follows. \Cref{sec:Setup} gives an overview of the Stopping and Trapping of Radioactive Isotopes for Precision Experiments (STRIPE) beamline, as well as new details on the two laser--trap configurations investigated in this work, together with the laser system, the photon collection region (PCR) and the RF amplifier. \Cref{sec:Meas} presents the results of this work: Doppler cooling and Coulomb-crystal formation (\cref{subsec:cooling}), the detection rate of the PCR (\cref{subsec:LCR}) and the trap storage lifetime (\cref{subsec:storage}), followed by spectroscopy of the $\Shalf\to\Dfivehalf$ clock transition in the between-electrodes and within-electrode configurations, from which the residual broadening mechanisms and the metastable $\Dfivehalf$-state lifetime are extracted (\cref{subsec:clock_spec}). \Cref{sec:Discussion} summarizes the performance of the setup and discusses the planned upgrades and future applications at the Ion Guide Isotope Separator On-Line (IGISOL) facility of the University of Jyv\"askyl\"a~\cite{MOORE2013208}, including in-trap laser spectroscopy of radioactive isotopes and cold-ion delivery to downstream precision experiments.

\section{\label{sec:Setup}EXPERIMENTAL SETUP}

\begin{figure*}[ht]
    \centering
    \includegraphics[width=0.95\linewidth]{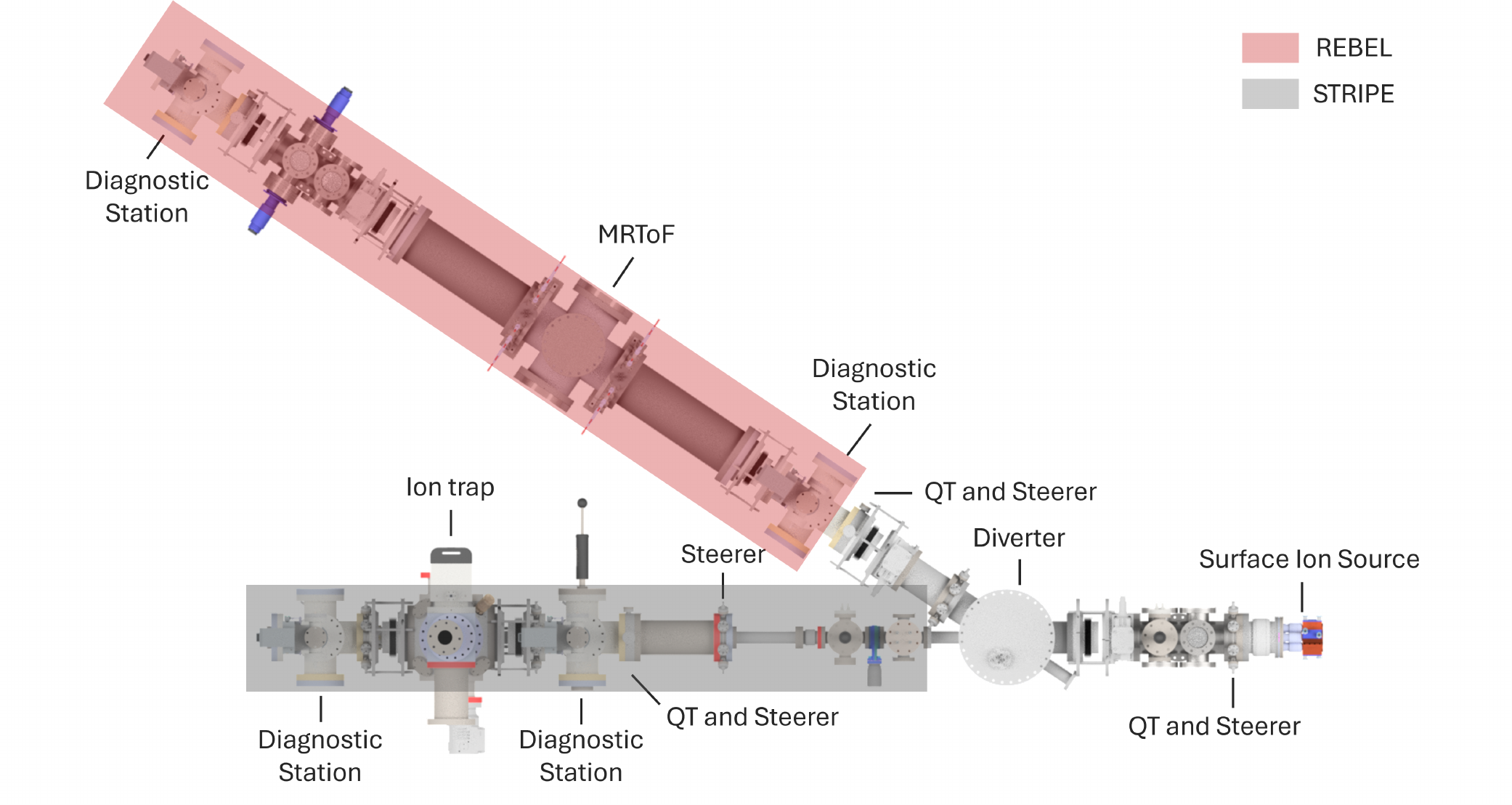}
    \caption{Overview of the offline laboratory commissioned at KU Leuven, where both the REBEL and STRIPE setups are present.}
    \label{fig:LAB_overview}
\end{figure*}

An overview of the beamline with both setups, the Resonant Excitation of Beams with Electromagnetic fields and Lasers (REBEL) setup and STRIPE, can be seen in \cref{fig:LAB_overview}, and a detailed description of its working principle, including ion production, beam transport and ion trapping, as well as first results, can be found in Ref.~\cite{Imgram2025}. In this overview, we focus on STRIPE and build on the previous work.

To deliver ions to the Paul trap, an ion source based on surface ionization was installed, in which natural Sr is heated inside a graphite crucible to create $\mathrm{Sr}^{+}$ ions, which then leave the source and are accelerated to beam energies up to \SI{10}{\kilo\electronvolt}. Along the beam path, sets of steerer and quadrupole triplet electrodes are used for steering and focusing.

\begin{figure*}[ht]
    \centering
    \includegraphics[width=0.95\linewidth]{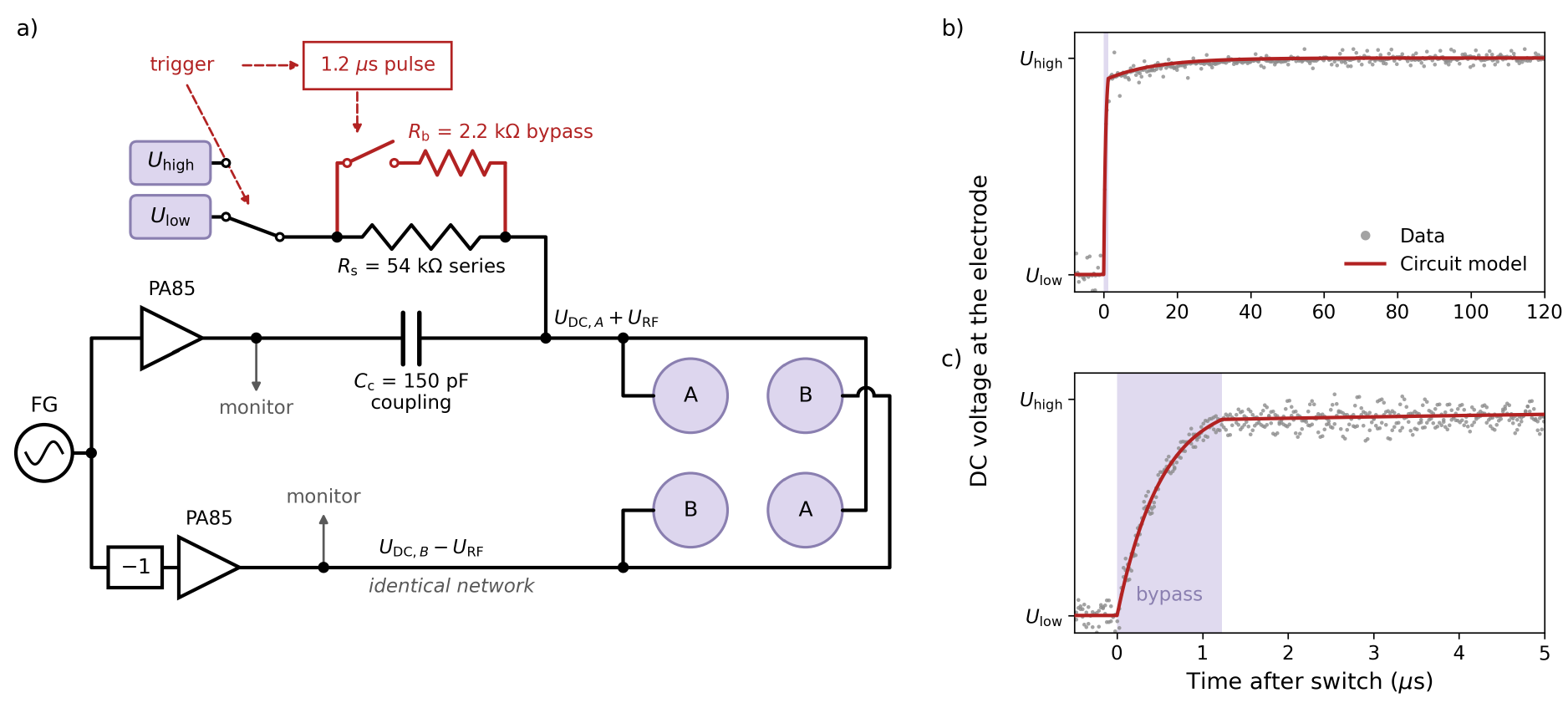}
    \caption{(a) Simplified schematic of the RF/DC unit for one trap segment. The DC network is drawn for the $+U_\mathrm{RF}$ phase (rods A); the $-U_\mathrm{RF}$ phase (rods B) has an identical one. The two rods of each diagonally opposite pair are wired together, and the bypass path is shown in red. (b) DC voltage at the electrode during a switch from $U_\mathrm{low}$ to $U_\mathrm{high}$, measured with the RF off on a monitor output, together with the circuit model (red). The shaded band marks the bypass pulse. (c) Zoom on the first \SI{5}{\micro\second}.}
    \label{fig:rf_dc_unit}
\end{figure*}

The ion trap itself is placed below the high voltage of the ion source, using a tunable DC offset between \SI{0}{\volt} and \SI{60}{\volt}. To minimize transmission losses as the ions are decelerated from \SI{10}{\kilo\electronvolt} down to a few \si{\electronvolt} once they enter the trap, a segmented decelerator is used to improve the beam properties during this stage. Radial confinement in the linear Paul trap is achieved by applying an RF field with typical frequencies $\Omega_{\mathrm{RF}}/2\pi$ in the range \SIrange{0.5}{1.7}{\mega\hertz} and a maximum applied peak-to-peak RF voltage of \SI{256}{\volt}, while the trap's segmentation allows the application of an additional DC potential to confine the ions in the axial direction. There are eight segments in total, each comprising four electrode rods in two pairs, carrying $U_{\mathrm{DC},A} + U_{\mathrm{RF}}$ and $U_{\mathrm{DC},B} - U_{\mathrm{RF}}$, respectively. The DC voltages $U_{\mathrm{DC},A}$ and $U_{\mathrm{DC},B}$ are individually adjustable between \SI{-24}{\volt} and \SI{24}{\volt}, and each can be switched between two preset values on an external trigger. This fast switching is what allows an incoming ion bunch to be captured in flight. Voltages are delivered by a custom-built unit, in which a single function generator signal and its inverted copy drive two PA85 power amplifiers, one per RF phase. The DC is coupled to the RF through an RC network, which allows operation over a \SI{2}{\mega\hertz} frequency range. To speed up the switching, the DC series resistor $R_\mathrm{s}$ is shorted for about \SI{1.2}{\micro\second} by a much smaller resistor $R_\mathrm{b}$. The electrode voltage therefore rises in two parts, a fast one while the bypass is closed and a slow one set by $R_\mathrm{s}$ after it opens, with a kink in between. \Cref{fig:rf_dc_unit} compares the measured voltage ramp with a simulation of the circuit. About \SI{90}{\percent} of the step is completed during the bypass pulse, and the voltage is within \SI{1}{\percent} of its new value after about \SI{40}{\micro\second}. The model includes a measured additional capacitance of about \SI{100}{\pico\farad} on the amplifier's output, of which roughly \SI{80}{\pico\farad} comes from the amplifier's load and the rest from the unit's own wiring and connector. A permanently smaller $R_\mathrm{s}$ would also make the switch faster, but it would load the amplifiers continuously, since the RF also drives a current through $R_\mathrm{s}$ into the DC supply. With $R_\mathrm{s} = R_\mathrm{b}$, this current would reach about \SI{60}{\milli\ampere} per channel at the maximum RF amplitude (\SI{0.5}{\ampere} in total), more than twice the amplifier's \SI{200}{\milli\ampere} rating, compared with about \SI{20}{\milli\ampere} in total for the same eight channels with $R_\mathrm{s} = \SI{54}{\kilo\ohm}$. The RF amplitude and frequency were monitored during operation through two dedicated outputs of the unit, one per RF phase.

\begin{figure*}[ht]
    \centering
    \includegraphics[width=0.8\linewidth]{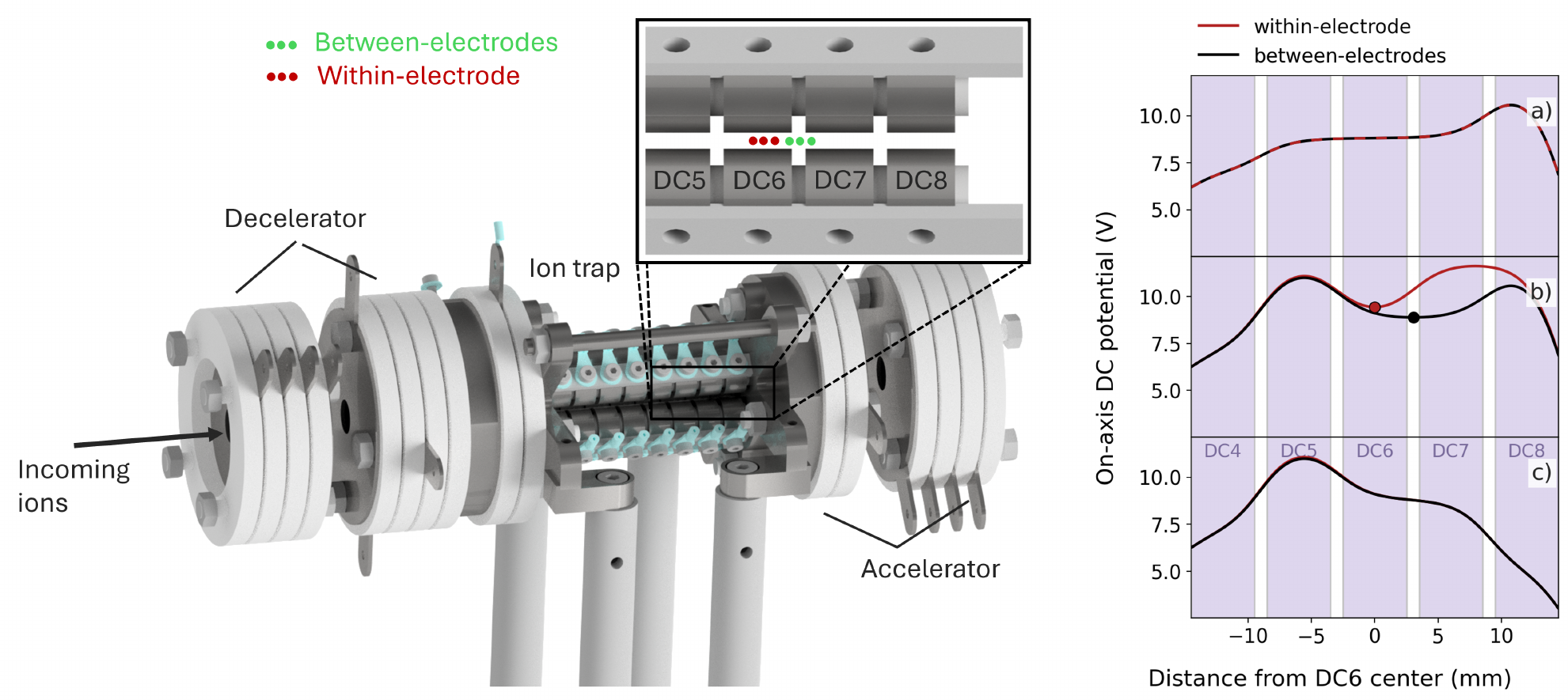}
    \caption{Rendering of the STRIPE linear Paul trap with the two ion-trapping configurations used in this work highlighted in the zoomed region: the between-electrodes configuration (ion crystal held between segments No.~6 and No.~7) and the within-electrode configuration (ion crystal held within segment No.~6). The DC potentials of these two configurations are illustrated on the right during (a) ion loading, (b) ion trapping and (c) ion extraction. The segment voltages shown are indicative; the same sequence works over a wide range of DC values.}
    \label{fig:linear_Paul_Trap}
\end{figure*}

An illustration of the linear Paul trap and the DC potential experienced by the ions during ion loading, trapping and extraction is given in \cref{fig:linear_Paul_Trap}. The electrodes are \SI{5}{\milli\metre} long and separated from their neighbors by a \SI{1}{\milli\metre} gap. The rods have a radius $r_{\mathrm{rod}} = \SI{4.6}{\milli\metre}$ and an inner-electrode half-distance of $r_{0} = \SI{3.0}{\milli\metre}$, resulting in $r_{\mathrm{rod}}/r_{0} \approx 1.53$. This is larger than the ideal value of 1.12 that best approximates a quadrupole field, as a result of a manufacturing fault~\cite{Imgram2025}. The segmentation allows the confinement and movement of an ion bunch along the axial direction by manipulation of the applied DC voltages. In this work, two specific trapping configurations were investigated in detail, henceforth defined as the \emph{between-electrodes} and \emph{within-electrode} configurations. In the former, the ion bunch is captured and overlapped with the laser beams between segments No.~6 and No.~7, while in the latter, the bunch is confined within segment No.~6, roughly at the center of the segment.

\begin{figure*}[ht]
    \centering
    \includegraphics[width=0.8\linewidth]{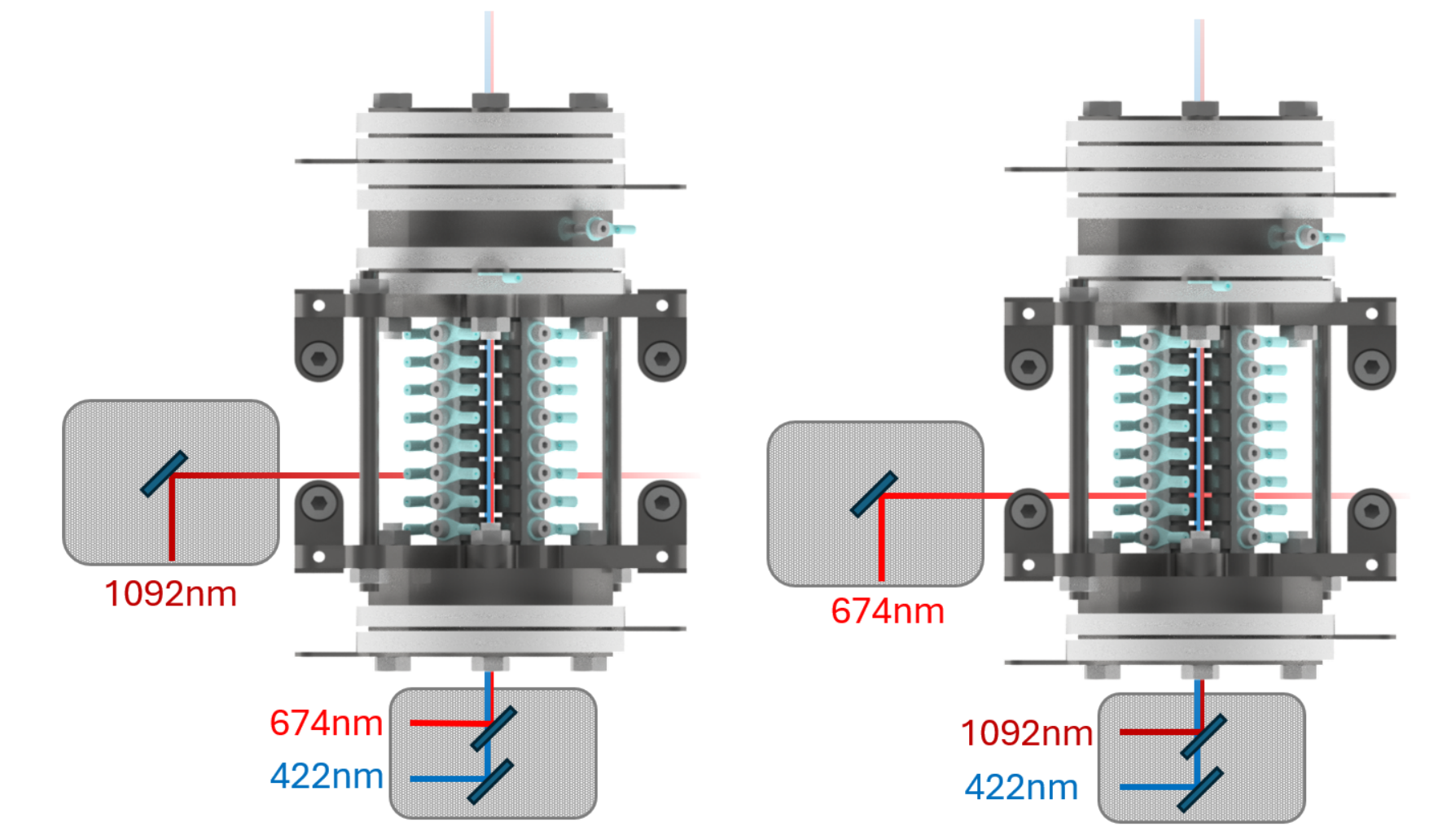}
    \caption{Overview of the laser setup used for Doppler laser cooling and in-trap laser spectroscopy for both the within-electrode (left) and between-electrodes (right) trapping configurations.}
    \label{fig:laser_paths}
\end{figure*}

In both configurations, the trapped-ion bunch was overlapped with two laser beams at \SI{422}{\nano\metre} and \SI{1092}{\nano\metre} for Doppler laser cooling, driving the $\Shalf \rightarrow \Phalf$ cooling and $\Dthreehalf \rightarrow \Phalf$ repump transitions in \SrIon{}. Since natural Sr contains other stable isotopes --- $^{84}$Sr, $^{86}$Sr and $^{87}$Sr, with representative natural abundances of \SI{0.56}{\percent}, \SI{9.86}{\percent} and \SI{7.00}{\percent} against \SI{82.58}{\percent} for $^{88}$Sr~\cite{Meija2016} --- which were co-trapped but not addressed by the cooling lasers, dark ions occasionally appeared in the Coulomb crystal. To improve the cooling efficiency by addressing a wider velocity class of ions, the \SI{422}{\nano\metre} laser light was sent through an acousto-optic modulator (AOM), from which both the $0^{\mathrm{th}}$ and the $-1^{\mathrm{st}}$ order, shifted \SI{248}{\mega\hertz} below the former, were recombined into the same path. A third laser was used to drive the dipole-forbidden \SI{674}{\nano\metre} $\Shalf \rightarrow \Dfivehalf$ transition. Since its natural linewidth is below \SI{1}{\hertz}, the observed width of this line is set entirely by broadening mechanisms such as Doppler broadening from the ion motion, micromotion and the laser linewidth, and therefore quantifies them directly. These laser beams were provided by three Toptica diode lasers (DL pro). The \SI{422}{\nano\metre} and \SI{1092}{\nano\metre} lasers were locked to a HighFinesse WS8-10 wavemeter, while the \SI{674}{\nano\metre} laser was free-running, its frequency read by the same wavemeter. The output of all lasers was fiber-coupled and transported to the two optical tables around the ion-trap setup, to allow for laser access from both the axial and radial directions, as illustrated in \cref{fig:laser_paths}.

During the measurements with these two different ion-trapping configurations, the three laser beams overlapped with the ions in two distinct ways, one for each configuration. In both cases, the \SI{422}{\nano\metre} cooling laser was injected axially, since the deceleration from \SI{10}{\kilo\electronvolt} leaves the ions' kinetic energy concentrated in the axial motion. Simulations of the in-flight capture of an ion bunch with properties similar to those in this work~\cite{PelonisMSc} show that a typical ion retains on the order of \SI{0.1}{\electronvolt} of kinetic energy axially, several times more than in either radial direction. Using a beam sampler, we attempted to direct part of the \SI{422}{\nano\metre} light along a radial direction, but no effect on the cooling process was seen. Trying to Doppler-cool with \SI{422}{\nano\metre} light solely from that direction was also unsuccessful, confirming that most of the energy to be extracted from the system was indeed along the axial direction. The \SI{674}{\nano\metre} clock laser and the \SI{1092}{\nano\metre} repump laser were, however, interchanged before entering the trap, with the \SI{674}{\nano\metre} and \SI{1092}{\nano\metre} lasers being sent axially/radially and radially/axially for the within-electrode/between-electrodes trapping configurations, respectively. The effect of the \SI{1092}{\nano\metre} repump was, as far as we could observe, similar from both directions. However, the change of its injection direction was necessary for improved alignment of the \SI{674}{\nano\metre} laser, because of the limited optical access. For the \SI{674}{\nano\metre} laser, both directions were explored to observe the dependence of the linewidth on the probing direction, as discussed in Ref.~\cite{Berkeland1998}.

\begin{figure}[h!]
    \centering
    \includegraphics[width=0.99\linewidth]{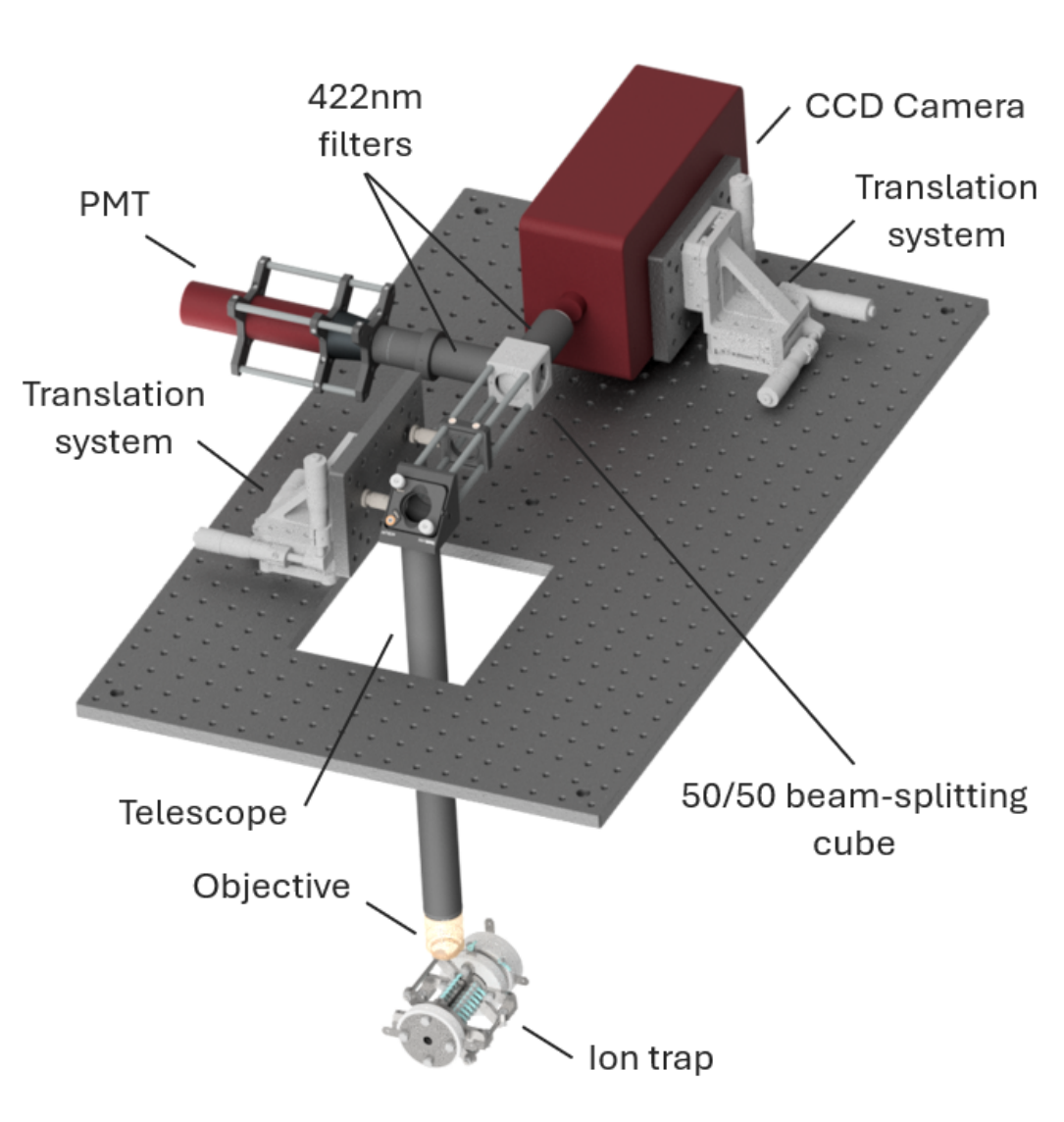}
    \caption{Overview of the photon collection region of STRIPE. A mirror and a
beamsplitter cube direct the photons collected by the objective toward
the PMT and camera.}
    \label{fig:LCR_modified}
\end{figure}

To detect the photons emitted from the ion--laser overlap region, a new photon collection region was designed and placed directly above the ion trap, as shown in~\cref{fig:LCR_modified}. The fluorescence is collected by a TL2X-SAP objective from Thorlabs, positioned \SI{56.3}{\milli\metre} above the ion position and mounted on a 3-axis translation stage. Light leaving the objective is deflected by \SI{90}{\degree} with a mirror and split 50/50 by a beamsplitter cube between a Hamamatsu H11123 PMT and an Andor iStar intensified CCD camera. A bandpass filter in each of the two arms transmits only light around \SI{422}{\nano\metre}.

\section{\label{sec:Meas}COMMISSIONING MEASUREMENTS}

Sr$^+$ ions delivered from the surface ion source at \SI{10}{\kilo\electronvolt} were decelerated to a few \si{\electronvolt} in the deceleration section and captured inside the Paul trap. The injected bunches were \SI{1}{\micro\second} long and initially contained several hundred ions. Depending on the trap's DC settings, bunches up to a few tens of \si{\micro\second} long could be captured. The overall efficiency from the ion source to crystallized ions was ${\sim}\SI{5}{\percent}$, comparable to the ${\sim}\SI{4}{\percent}$ from source to laser-cooled ions reported in Ref.~\cite{Imgram2025}. The ion trap was floated \SI{8}{\volt} below the ion source. It was driven at $\Omega_{\mathrm{RF}}/2\pi = \SI{1.6}{\mega\hertz}$ (\SI{1.2}{\mega\hertz} for the crystal-formation sequence of \cref{fig:ion_crystal_formation}) with an RF amplitude of \SI{70}{\volt} peak-to-peak at the electrodes, corresponding to a radial secular frequency of order \SI{100}{\kilo\hertz}. The segmentation of the trap was used to create a DC slope of roughly \SI{2}{\volt\per\centi\metre} to slow down the ions and capture them. Doppler laser cooling of $^{88}$Sr$^+$ was then performed by directing the dual-mode \SI{422}{\nano\metre} laser (\SI{3.0}{\milli\watt} of $0^{\mathrm{th}}$-order light at roughly \SI{60}{\mega\hertz} and \SI{3.1}{\milli\watt} of $-1^{\mathrm{st}}$-order light at roughly \SI{310}{\mega\hertz} below resonance) and the \SI{1092}{\nano\metre} repump (\SI{340}{\micro\watt} at roughly \SI{20}{\mega\hertz} below resonance) through the trap volume. These detunings are uncertain at the \SI{10}{\mega\hertz} level, since the wavemeter was not calibrated against an absolute frequency reference and the laboratory was not temperature-stabilized.

Most measurements discussed in this work were performed in the within-electrode trapping and laser-cooling configuration (see \cref{fig:linear_Paul_Trap} and \cref{fig:laser_paths}). The exceptions are the Coulomb-crystal formation sequence of \cref{subsec:cooling} and the spectroscopy of \cref{subsubsec:config1}, which were performed in the between-electrodes one. During the measurements on the $\Shalf\to\Dfivehalf$ transition, \SI{320}{\micro\watt} of \SI{674}{\nano\metre} light was focused inside the trap.

\subsection{\label{subsec:cooling}Doppler cooling and Coulomb-crystal formation}

\begin{figure*}[ht]
    \centering
    \includegraphics[width=0.95\linewidth]{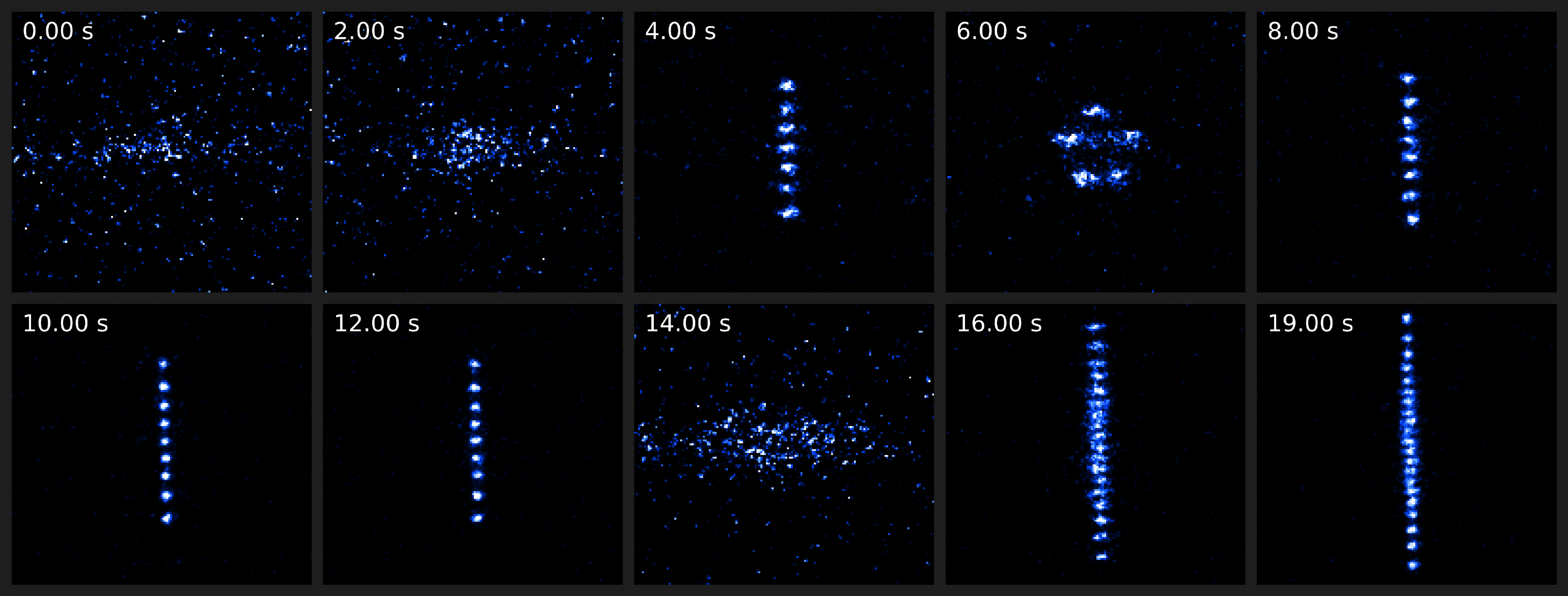}
    \caption{Camera frames at several times $t$ after an ion bunch entered the trap, illustrating the formation of a Coulomb crystal of \SrIon{} ions. At $t=\SI{14}{\second}$, the trapping well was briefly lowered to allow the injection of a second ion bunch.}
    \label{fig:ion_crystal_formation}
\end{figure*}

After the ion bunches had been loaded following the sequence described above, the ions formed a well-defined Coulomb crystal within several seconds of continuous laser cooling. A typical example is depicted in~\cref{fig:ion_crystal_formation}, taken using the CCD camera (\SI{300}{\milli\second} exposure). The trap also supports faster cooling of a second injected bunch by a pre-cooled crystal. This is shown in the same \cref{fig:ion_crystal_formation}, where, at $t=\SI{14}{\second}$, the trapping well was lowered for roughly \SI{15}{\micro\second} to allow the injection of a second Sr$^+$ bunch. Unlike the first one, which required several seconds to form a Coulomb crystal, the second experienced faster crystallization in roughly \SI{2}{\second}. This can in principle also be applied to different isotopes and elements, which would further shorten the effective cooling timescale for ions of interest, including ones that are not directly laser-coolable. It is worth noting that the Coulomb crystal of 7 ions at $t=\SI{4}{\second}$ grew to 8 ions at $t=\SI{8}{\second}$ and even 9 at $t=\SI{9}{\second}$. This behavior was observed regularly with our setup and is consistent with ions with higher-than-average energy, orbiting on larger trajectories, joining the crystal once they were finally cooled close to the Doppler limit.

Fluorescence is already visible in the first \SI{300}{\milli\second} exposure of~\cref{fig:ion_crystal_formation}, showing that within a few hundred \si{\milli\second} the ions had been slowed into resonance with the \SI{422}{\nano\metre} laser, marking the onset of cooling. Similar behavior was observed with the ToF spectra reported in Ref.~\cite{Imgram2025}. Unlike the present work, where laser-cooling modulation was achieved with an AOM, the results from Ref.~\cite{Imgram2025} were based on a frequency-swept \SI{422}{\nano\metre} cooling laser. There, a substantially narrowed ToF peak appeared after ${\sim}\SI{250}{\milli\second}$, compared with ${\sim}\SI{1}{\second}$ when no sweeping was applied. Both laser modulation schemes extend efficient cooling to a wider range of ion velocities than a single fixed-detuning tone and open the way to the application of in-trap spectroscopy to short-lived isotopes with half-lives of a few hundred \si{\milli\second}. The overlap with the $-1^{\mathrm{st}}$-order beam proved essential even after initial crystallization. Without it, already-cold crystals were observed to melt from time to time, most likely after heating by collisions with the background gas.

The appearance of Coulomb crystals places an upper limit on the temperature of the ions. Crystallization requires the Coulomb interaction energy between neighboring ions to exceed their thermal energy by a factor of about 170, which in ion traps corresponds to temperatures of order \SI{10}{\milli\kelvin} or below~\cite{Drewsen2015}. We therefore estimate the temperature of our crystallized ions to be ${\sim}\SI{10}{\milli\kelvin}$.

\subsection{\label{subsec:LCR}Detection rate of the photon collection region}

A three-dimensional Coulomb crystal of order 50 \SrIon{} ions was loaded and held under continuous Doppler cooling for \SI{25}{\minute}, with the PMT continuously recording the fluorescence count rate. At several time points during three independent measurements, CCD camera images of the crystal were acquired and the number of ions $N_{\mathrm{ion}}$ was counted by visual inspection. Larger uncertainties were assigned to images in which individual ions could not be cleanly resolved. The background contribution from scattered \SI{422}{\nano\metre} laser light was determined by recording the PMT signal immediately after ejecting all ions from the trap and was subtracted from all readings.

\begin{figure}[h!]
    \centering
    \includegraphics[width=\linewidth]{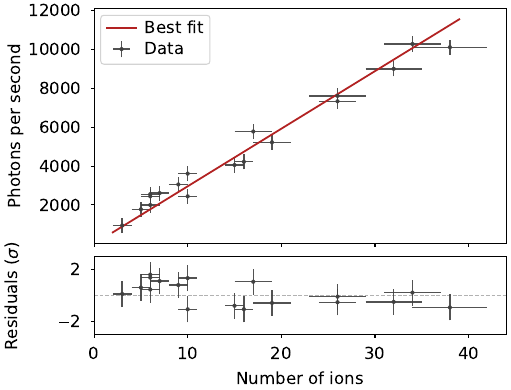}
    \caption{Background-subtracted PMT count rate as a function of the number of trapped \SrIon{} ions counted on concurrent camera images. The solid line is a linear fit through the origin yielding $\eta_\mathrm{PCR} = \SI{296(9)}{photons\,\second^{-1}\,ion^{-1}}$ ($\chi^{2}_{\nu} = 0.84$). Horizontal error bars reflect the visual ion-counting uncertainty.}
    \label{fig:LCR_efficiency}
\end{figure}

A linear model $R = \eta_\mathrm{PCR} \cdot N_{\mathrm{ion}}$ was fitted to the
background-subtracted count rate as a function of $N_{\mathrm{ion}}$
(\cref{fig:LCR_efficiency}), where $R$ is the photon count rate, with the uncertainty in $N_{\mathrm{ion}}$ propagated through the model into the weight of each point. The fit yielded $\eta_\mathrm{PCR} = \SI{296(9)}{photons\,\second^{-1}\,ion^{-1}}$ recorded with the PMT, with roughly the same rate sent to the camera.

This measured rate can be compared with the rate expected per ion at the PMT,
\begin{equation}
  R_\mathrm{th} = R_\mathrm{sc}\,\frac{\Omega}{4\pi}\,
                  T_\mathrm{v}\,R_\mathrm{m}\,T_\mathrm{BS}\,T_\mathrm{f}\,
                  \eta_\mathrm{PMT},
  \label{eq:Rth}
\end{equation}
where $R_\mathrm{sc}$ is the \SI{422}{\nano\metre} photon rate scattered per ion into $4\pi$, $\Omega/4\pi=\SI{0.251}{\percent}$ the fraction of the solid angle collected by the objective for isotropic emission, obtained from ray-tracing simulations of the setup, $T_\mathrm{v}\approx0.92$ the transmission of the uncoated viewport, $R_\mathrm{m}\approx0.995$ the mirror reflectivity, $T_\mathrm{BS}=0.50$ the beamsplitter transmission into one arm, $T_\mathrm{f}\approx0.90$ the bandpass-filter transmission and $\eta_\mathrm{PMT}\approx0.21$ the effective PMT detection efficiency at \SI{422}{\nano\metre}. Together, these six factors give a detection probability of about \num{2e-4} per emitted photon.

The scattering rate $R_\mathrm{sc}$ is estimated from a simplified three-level rate-equation model of the \SrIon{} cooling cycle, detailed in Appendix~\ref{app:lcr_model}. Since the laser beam diameters at the ion position could not be measured \emph{in situ}, we estimate a $1/e^{2}$ beam diameter of roughly \SI{4}{\milli\metre}. With the laser powers and detunings given above, the model gives $R_\mathrm{sc}\approx\SI{2.6e6}{photons\,\second^{-1}\,ion^{-1}}$, and \cref{eq:Rth} then predicts $R_\mathrm{th}\approx\SI{570}{photons\,\second^{-1}\,ion^{-1}}$, about twice the measured $\eta_\mathrm{PCR}$.

\begin{figure}[h!]
    \centering
    \includegraphics[width=\linewidth]{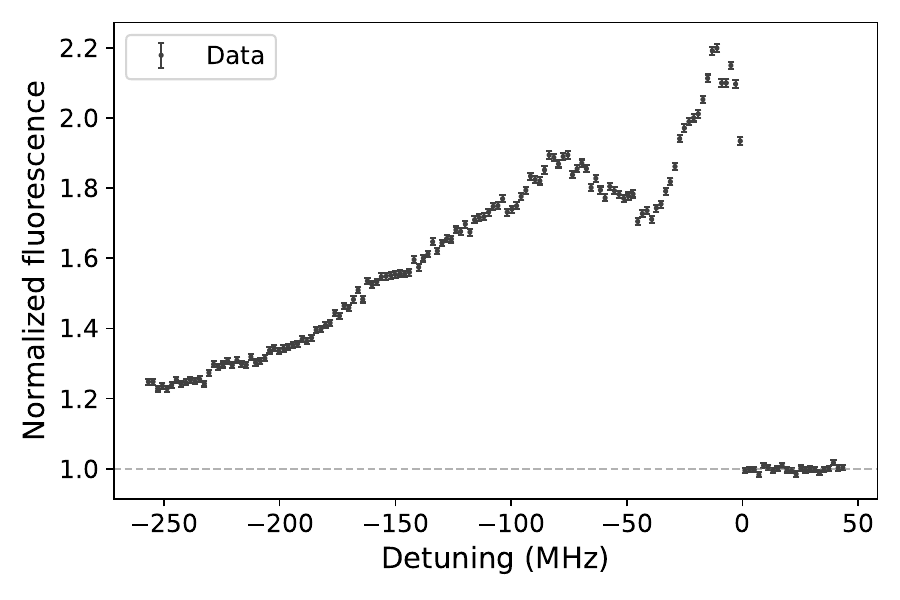}
    \caption{Typical scan of the $\Shalf\rightarrow\Phalf$ cooling transition, normalized to the background of scattered laser light. The dark resonance is visible as a dip roughly \SI{40}{\mega\hertz} below resonance.}
    \label{fig:cooling_line_scan}
\end{figure}

A typical scan of the cooling transition is shown in \cref{fig:cooling_line_scan}, taken in the same configuration. Close to resonance the laser starts to heat the ions and the fluorescence drops abruptly to the background, so the blue side of the line is not recorded. The scan also shows a dark resonance of the $\Lambda$-type system formed by the \SI{422}{\nano\metre} and \SI{1092}{\nano\metre} lasers. Its position and depth depend on the intensities and polarizations of the cooling and repump lasers, as well as the magnetic field, none of which were precisely controlled. For this scan, the repump was locked roughly \SI{40}{\mega\hertz} below resonance. Such dark states are not included in the model and lower the scattering rate, which may account for part of the difference between $R_\mathrm{th}$ and the measured $\eta_\mathrm{PCR}$, together with any wavemeter drift.

The detection rate is limited both by the small solid angle and by the fluorescence rate per ion. The former is fixed by the geometry of the present trap, while the latter is set by the \SI{422}{\nano\metre} laser parameters. The large detunings required to address the incoming hot ions keep the $\Phalf$ population at the percent level, well below the ${\sim}\SI{33}{\percent}$ of a saturated cooling cycle. Increasing the \SI{422}{\nano\metre} intensity at the ion position would raise the detection rate without any change to the photon collection region itself.

\subsection{\label{subsec:storage}Storage lifetime}

\begin{figure}[h!]
    \centering
    \includegraphics[width=\linewidth]{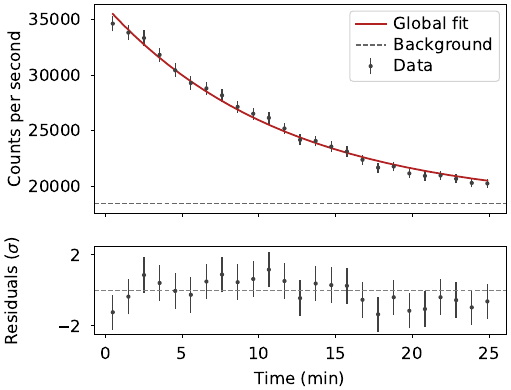}
    \caption{Ion storage-lifetime measurement: decay of the fluorescence count rate over a single cycle, with the global fit of \cref{eq:storage_decay} to all three cycles. The curve levels off at the constant background $R_{\mathrm{bkg}}$ (dashed line).}
    \label{fig:trapped_ion_lifetime}
\end{figure}

The datasets used to determine $\eta_\mathrm{PCR}$ above were also analyzed to extract the trap storage lifetime. The PMT count rate was recorded over the full \SI{25}{\minute} run and fitted with an exponential decay on a constant background,
\begin{equation}
    R(t) = R_{0}\,e^{-t/\tau_{\mathrm{trap}}} + R_{\mathrm{bkg}}.
    \label{eq:storage_decay}
\end{equation}
Three independent measurement cycles, one of which is shown in \cref{fig:trapped_ion_lifetime}, were fitted together with a shared $\tau_{\mathrm{trap}}$. The offset $R_{\mathrm{bkg}}$ was fixed to the rate of \SI{1.841(3)e4}{counts\,\second^{-1}} measured in a dedicated run with the \SI{422}{\nano\metre} light on and no ions in the trap, since $R_{\mathrm{bkg}}$ and $\tau_{\mathrm{trap}}$ are strongly anti-correlated in the fit (correlation coefficient $-0.96$).

The global fit yielded a storage lifetime of $\tau_{\mathrm{trap}} = \SI{11.6(10)}{\minute}$. The uncertainty is the statistical limit set by the number of ions lost during the measurement, $\tau_{\mathrm{trap}}/\sqrt{N}$ with $N \approx 146$ over the three cycles, which exceeds the uncertainty returned by the fit. This is more than an order of magnitude longer than the laser-cooled storage time of ${\sim}\SI{20}{\second}$ reported in Ref.~\cite{Imgram2025}. We attribute this to more effective Doppler cooling, since adding the $-1^{\mathrm{st}}$-order cooling mode helped prevent the Coulomb crystals from melting, potentially through ion--gas collisions. Such a collision with the background gas can melt the Coulomb crystal, after which RF heating either ejects ions from the trap or leaves them in orbits too hot for the cooling laser to recapture~\cite{Pagano2019}. The present limitation is likely the chamber pressure of approximately \SI{2e-8}{\milli\bar}. Isolating the trap section from the rest of the beamline was separately observed to bring the pressure below \SI{1e-9}{\milli\bar} with the same pump configuration, showing that the residual gas load originates upstream, mainly in the surface ion-source chamber. That chamber sat at \SI{3e-8}{\milli\bar}, set by its elastomer-sealed KF100 flange, which is suitable only down to the low \SI{1e-8}{\milli\bar} range. The pressure at the trap was the same whether or not the ion source was being operated.

\subsection{\label{subsec:clock_spec}E2-transition spectroscopy and
\texorpdfstring{$\Dfivehalf$}{4d 2D5/2} state lifetime}

We performed spectroscopy of the \mbox{$\Shalf \to \Dfivehalf$} electric-quadrupole clock transition at \SI{674}{\nano\metre} in two distinct ion-trapping configurations within the trap to investigate possible broadening mechanisms: the between-electrodes configuration, in which the ion crystal is held between segments No.~6 and No.~7, and the within-electrode configuration, in which it is held within segment No.~6 (see~\cref{fig:laser_paths}). In both subsections below, three-dimensional ion crystals of order 50 ions were loaded.

The \SI{422}{\nano\metre} cooling and \SI{1092}{\nano\metre} repump lasers overlapped with the ions continuously throughout every scan, both locked to the wavemeter, while the \SI{674}{\nano\metre} laser was scanned across the resonance. In the within-electrode configuration the \SI{674}{\nano\metre} laser was stepped in \SI{1}{\mega\hertz} increments across \SI{140}{\mega\hertz}. The broader between-electrodes resonance was covered in \SI{10}{\mega\hertz} increments across \SI{450}{\mega\hertz}.

\subsubsection{\label{subsubsec:config1}Between-electrodes configuration}

\begin{figure}[h!]
    \centering
    \includegraphics[width=\linewidth]{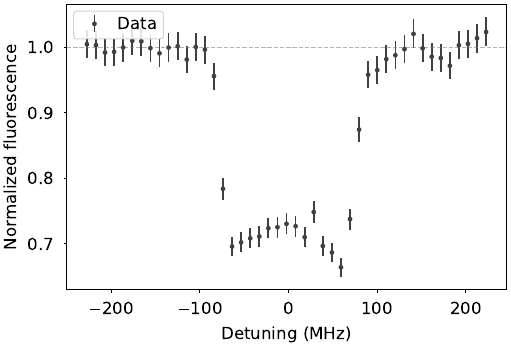}
    \caption{Resonance spectrum of the $\Shalf \rightarrow \Dfivehalf$ transition, recorded in the between-electrodes configuration. The dips at the edges of the resonance are characteristic of considerable micromotion.}
    \label{fig:674nm_resonance_radially}
\end{figure}

In the between-electrodes configuration, the ions were confined between segments No.~6 and No.~7, and the \SI{422}{\nano\metre} and \SI{1092}{\nano\metre} lasers were sent axially through the trap, while the \SI{674}{\nano\metre} clock laser was sent radially. A summed spectrum from the measurements on the $\Shalf \rightarrow \Dfivehalf$ resonance is depicted in~\cref{fig:674nm_resonance_radially}. Unlike in typical excitation spectra, the resonance appears as a drop in photon count rate, since ions on resonance with the transition were shelved to the $\Dfivehalf$ level and stopped fluorescing. The count rate was also normalized to a fitted linear background, to account for the decay rate of the trap established in \cref{subsec:storage}.

This resonance exhibits a broad lineshape with a flat top and two enhanced dips at the edges of the line. Following the discussion from Berkeland \textit{et al.}~\cite{Berkeland1998}, an ion displaced from the nodal line of the trap's RF field by a residual static field, or driven by a small phase mismatch between the RF electrodes, acquires a sinusoidal velocity component $\mathbf{v}_{m}(t)=\mathbf{v}_{m}\cos(\Omega_{\mathrm{RF}} t)$, where $\Omega_{\mathrm{RF}}$ is the trap's angular drive frequency. In the rest frame of such an ion, the spectroscopy laser is first-order Doppler modulated with the dimensionless index
\begin{equation}
\xi \;=\; \frac{|\mathbf{k}\!\cdot\!\mathbf{v}_{m}|}{\Omega_{\mathrm{RF}}}
\;=\; \frac{k\,v_{\max}}{\Omega_{\mathrm{RF}}},
\label{eq:micromotion_beta}
\end{equation}
where $v_{\max}=|\hat{\mathbf{k}}\cdot\mathbf{v}_{m}|$ is the peak micromotion velocity along
the laser wave vector $\mathbf{k}$. In the slow-modulation regime relevant here, with $\Omega_{\mathrm{RF}}/2\pi = \SI{1.6}{\mega\hertz}$ much smaller than the Gaussian width of the measured profile, the micromotion sidebands described by Berkeland \textit{et al.}~\cite{Berkeland1998} merge into a continuous profile whose half-extent is set by the peak Doppler excursion $k v_{\max}$. For large $\xi$ this profile develops two maxima near $\pm k v_{\max}$ and a depression at the line center, as observed in~\cref{fig:674nm_resonance_radially}, indicating that the ions were in the large-$\xi$ regime.

\subsubsection{\label{subsubsec:config2}Within-electrode configuration: Zeeman-resolved spectroscopy}

\begin{figure}[h!]
    \centering
    \includegraphics[width=\linewidth]{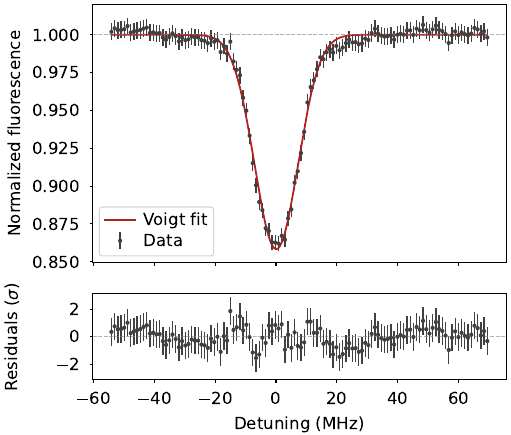}
    \caption{Resonance spectrum of the $\Shalf \rightarrow \Dfivehalf$ transition, recorded in the within-electrode configuration.}
    \label{fig:scan_sum_0A_shifted_fit}
\end{figure}

In the within-electrode configuration, the ions were confined within segment No.~6, with the \SI{422}{\nano\metre} and \SI{674}{\nano\metre} lasers sent axially through the trap and the \SI{1092}{\nano\metre} repump sent radially. The resulting resonances (see~\cref{fig:scan_sum_0A_shifted_fit}) were substantially narrower than in the between-electrodes configuration and showed none of the edge dips, indicating that the micromotion-induced broadening along the laser direction was strongly reduced. This may be due to the trapping position, the direction of the probe laser or a combination of the two.

In an ideal linear (2D) Paul trap, an ion undergoes micromotion only along directions in which the RF field contributes to the confinement, the radial ones in this case. Intrinsic micromotion, arising from the coupling of the secular motion to the radial RF field, is always present, and a stray DC field that displaces the ion equilibrium from the RF null adds a further excess micromotion of static amplitude --- both purely radial. The axial direction is confined electrostatically and, in an ideal trap, carries no RF field and therefore no micromotion. In a segmented trap this no longer holds strictly: geometric imperfections and RF pickup onto axially adjacent electrodes with unequal amplitude or phase leave a residual axial RF field, and hence axial micromotion.

Since both the intrinsic and the DC-induced excess micromotion are radial, an axially propagating \SI{674}{\nano\metre} laser is insensitive to them at first order. Any residual micromotion broadening of the $\Shalf\to\Dfivehalf$ line must then originate either from an axial RF-field component or from the radial component projected onto an imperfectly aligned probe beam. The narrowing is therefore attributed primarily to the probe geometry, which removes the dominant radial micromotion from the measurement. Whether the trapping position contributes in addition cannot be established from these data. An ion close to the center of an electrode is, however, expected to couple only weakly to that electrode's residual axial RF field, so that a within-electrode position should be less sensitive to per-electrode imbalance than one between electrodes.

In~\cref{fig:scan_sum_0A_shifted_fit}, the obtained spectrum was fitted with a simple Voigt model on top of a linear background, which yielded $\FWHMG = \SI{16.8(9)}{\mega\hertz}$. To resolve any underlying Zeeman structure due to the presence of a stray magnetic field, a single pair of Helmholtz-like coils was installed around the trap and further scans were recorded under a sequence of coil currents $I \in \{1, 2, 3\}\,\si{\ampere}$. The axis of the coils and the resulting induced magnetic field were perpendicular to the \SI{422}{\nano\metre} laser beam.

The polarization of the \SI{674}{\nano\metre} beam was measured with a Schäfter+Kirchhoff SK010PA polarization analyzer and found to be linear with an angle of \SI{77.2}{\degree} between the electric-field vector and the coil-induced magnetic-field direction. Stray-field contributions outside the trap chamber were estimated by an external Hall-probe survey and an upper limit of \SI{0.5}{\gauss} inside the trap volume was established. The expected coil-induced field at the ions exceeded this stray-field value for all coil measurements.

In the presence of $\vb{B}$, the $\Shalf \to \Dfivehalf$ transition splits into ten Zeeman components $\ket{\Shalf, m_{S}} \to \ket{\Dfivehalf, m_{D}}$ ($\Delta m \in \{0,\pm1,\pm2\}$), shifted from the field-free center $\nu_{0}$ by $\Delta\nu(m_{S}, m_{D}) = (\mu_{B}B/h)(g_{D}m_{D}-g_{S}m_{S})$, with Land\'{e} factors $g_{S} = 2.002256$ and $g_{D} = 1.200334$. In the absence of measured values for \SrIon{}, we assume those of the homologous Ca$^{+}$ levels, $g_{S}$ from Tommaseo \textit{et al.}~\cite{Tommaseo2003} and $g_{D}$ from Chwalla \textit{et al.}~\cite{Chwalla2009}. The relative amplitude of each component is the product of an atomic factor (a squared Wigner $3j$ symbol) and a geometric factor $W_{q}$ derived from the rank-2 field tensor coupling the laser polarization and propagation direction to the quantization axis defined by $\vb{B}$. The full derivation is presented in Appendix~\ref{app:zeeman}. The model takes the quantization axis along the coil field, which is perpendicular to $\hat{k}_{674}$. For this geometry $W_{0} = 0$ exactly, regardless of the polarization state, suppressing all $\Delta m = 0$ transitions; the $|\Delta m| = 2$ geometric weights carry ${\sim}\SI{95}{\percent}$ of $\sum_{q} W_{q}$ and the $|\Delta m| = 1$ the remaining ${\sim}\SI{5}{\percent}$. The quantization axis is, however, set by the total field at the ions, which also contains the stray field of unknown direction. The sensitivity of the fit to this assumption is evaluated below. Each component is broadened into a Voigt profile with shared Gaussian FWHM $\FWHMG$. The Lorentzian width $\Gamma_{L}$ converged to values compatible with zero during initial fits and was subsequently fixed to zero in the final fit.

\begin{figure*}[t]
    \centering
    \includegraphics[width=\textwidth]{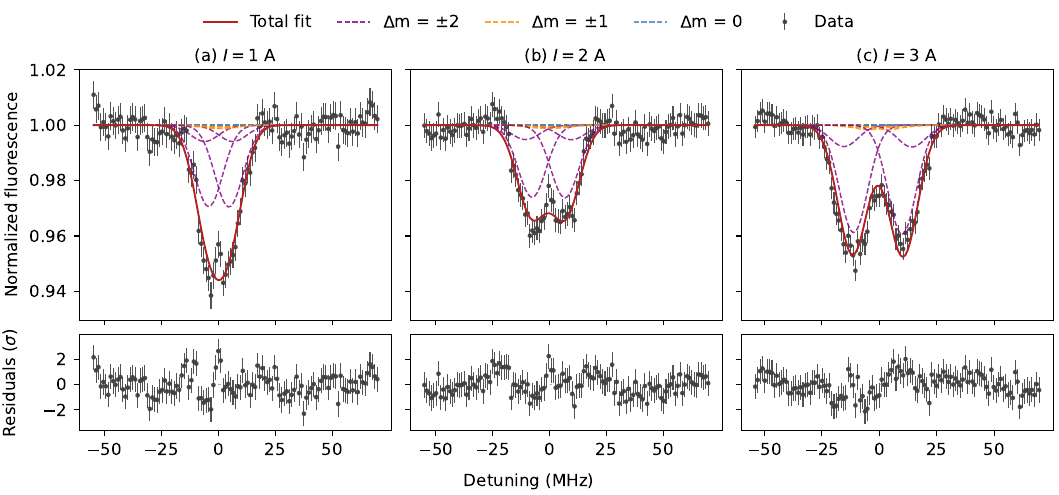}
    \caption{Zeeman-resolved spectra of the $\Shalf \to \Dfivehalf$ clock transition at coil currents of (a) \SI{1}{\ampere}, (b) \SI{2}{\ampere} and (c) \SI{3}{\ampere} (within-electrode configuration). The dashed colored lines show the individual $\Delta m$ contributions of the global fit; the lower panels show the residuals.}
    \label{fig:Zeeman_panels}
\end{figure*}

The combined spectra at $I \in \{1, 2, 3\}\,\si{\ampere}$ (\cref{fig:Zeeman_panels}) were fitted simultaneously with a shared $\FWHMG$ and individual magnetic-field magnitudes. The fit yielded $\FWHMG = \SI{14.5(8)}{\mega\hertz}$ together with one magnetic-field magnitude per coil current. The uncertainty on $\FWHMG$ includes the scan-to-scan scatter of the individual spectra. The correlations between $\FWHMG$ and the field magnitudes are small, since the field sets the spacing of the Zeeman components and $\FWHMG$ their individual width. The fitted magnitudes scale linearly with the coil current, at ${\approx}\SI{1.1}{\gauss\per\ampere}$. Neither the polarization angle, varied by $\pm\SI{5}{\degree}$ to cover the analyzer uncertainty and possible birefringence of the vacuum window, nor the unknown direction of a stray field of up to \SI{0.5}{\gauss}, tested by refitting with the quantization axis along the total field, changes $\FWHMG$ by more than a few tenths of a \si{\mega\hertz}, well within its uncertainty. This Zeeman-resolved $\FWHMG$ is smaller than the value obtained from the Voigt fit of the spectrum recorded without the coil-induced field. The difference, about \SI{2}{\mega\hertz}, amounts to about two standard deviations, and a stray field within the stated limit accounts for at most \SI{0.4}{\mega\hertz} of it.

The natural linewidth of this transition is \SI{0.407(2)}{\hertz}, from the measured $\Dfivehalf$ lifetime of \SI{0.3908(16)}{\second}~\cite{Letchumanan2005} discussed in \cref{subsubsec:lifetime_D52}, and no effect of power broadening from the \SI{674}{\nano\metre} laser was observed in either configuration. This was verified by recording the linewidth at different laser powers. With laser linewidths below \SI{200}{\kilo\hertz} and a wavemeter resolution of \SI{150}{\kilo\hertz}, technical broadening contributes at most \SI{0.3}{\mega\hertz} in quadrature. Photon scattering on the continuously applied cooling transition adds a Lorentzian contribution of \SIrange{0.2}{0.4}{\mega\hertz}. Thermal Doppler broadening from residual secular motion contributes a Gaussian FWHM of at most \SI{3.4}{\mega\hertz}, evaluated at the ${\sim}\SI{10}{\milli\kelvin}$ temperature estimate of \cref{subsec:cooling}. Together, these identified contributions account for only a few \si{\mega\hertz} of the measured ${\approx}\SI{15}{\mega\hertz}$, leaving a broadening of order \SI{10}{\mega\hertz} that we attribute mainly to residual micromotion. The latter is significantly reduced in this trapping geometry compared with the between-electrodes configuration.

\subsubsection{\label{subsubsec:lifetime_D52}\texorpdfstring{$\Dfivehalf$}{4d 2D5/2} metastable lifetime}

The measurement of the $\Dfivehalf$ level lifetime was performed in the within-electrode configuration, with each scan consisting of 100 successive \SI{3}{\second} cycles. In each cycle the \SI{674}{\nano\metre} laser was sent in for \SI{100}{\milli\second} via a mechanical shutter, then blocked for \SI{1.9}{\second}, followed by \SI{1}{\second} of dead time before the next cycle. This \SI{2.9}{\second} blocking period also ensured that almost all ions shelved to the $\Dfivehalf$ level had decayed before the next cycle. Both the \SI{422}{\nano\metre} cooling and the \SI{1092}{\nano\metre} repump lasers were sent into the trap during the whole measurement. The PMT counts recorded in the \SI{1.9}{\second} dark window were summed over the 100 cycles of each scan, and the resulting count rate $N(t)$ was fitted with
\begin{equation}
  N(t) = C_{0} - A_{0}\,e^{-t/\tau},
  \label{eq:lifetime_model}
\end{equation}
where $\tau$ is the $\Dfivehalf$ lifetime, $C_{0}$ the steady-state fluorescence level and $A_{0}$ the recovery amplitude proportional to the number of shelved ions. The earliest fit time $t_{c}$ in each scan was selected by the chi-squared saturation method, following the approach of Ref.~\cite{PhysRevA.110.L010802}. The fit was repeated for a series of trial $t_{c}$ values and the cutoff was chosen as the earliest $t_{c}$ at which the reduced $\chi^{2}_{\nu}$ entered its saturation plateau, with a per-scan systematic uncertainty assigned from the spread of $\tau$ values across the plateau. The mechanical shutter needed ${\sim}\SI{3}{\milli\second}$ to change state, so the earliest fit time was restricted to $t_{c}\geq\SI{103}{\milli\second}$.

\begin{figure}[h]
    \centering
    \begin{subfigure}[t]{\linewidth}
        \centering
        \includegraphics[width=\linewidth]{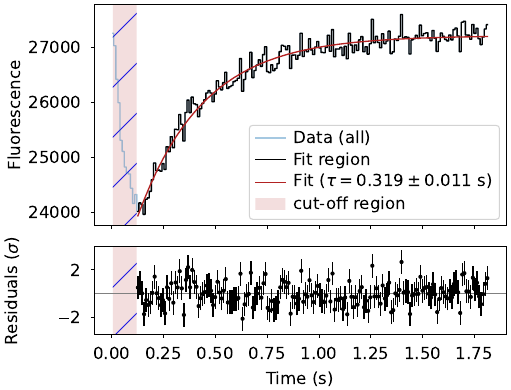}
        \caption{Scan 1, summed over its 100 cycles, with the fit of \cref{eq:lifetime_model} above the cutoff $t_{c}$ (shaded).}
        \label{fig:lifetime_scan382_hist}
    \end{subfigure}\\[1ex]
    \begin{subfigure}[t]{\linewidth}
        \centering
        \includegraphics[width=\linewidth]{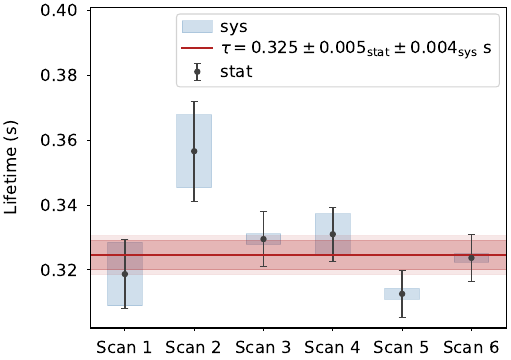}
        \caption{Lifetimes of the six scans (error bars: statistical; boxes: systematic) and their weighted mean, \cref{eq:lifetime_weighted}.}
        \label{fig:674nm_lifetime_avg}
    \end{subfigure}
    \caption{Metastable $\Dfivehalf$ lifetime measurement.}
    \label{fig:674nm_lifetime}
\end{figure}

Six independent scans were analyzed. The fit of the first is shown in \cref{fig:lifetime_scan382_hist}, and the lifetimes of all six in \cref{fig:674nm_lifetime_avg}. The inverse-variance-weighted mean over the six scans, with the statistical uncertainty taken as the larger of the internal and external estimates following the standard Birge-ratio convention~\cite{Birge1932}, gives
\begin{equation}
\tau_{\mathrm{obs}} = 0.325(5)_{\mathrm{stat}}(4)_{\mathrm{sys}}\,\si{\second}.\label{eq:lifetime_weighted}
\end{equation}
The per-scan systematic uncertainty was treated as fully correlated across scans and combined as a weighted linear sum,
\begin{equation}
  \sigma_{\mathrm{sys}} = \sum_{i}(w_{i}/W)\,\sigma_{\mathrm{sys},i},\qquad W = \sum_{i}w_{i},
\end{equation}
with $w_{i}$ the statistical-only weights.

The observed lifetime $\tau_{\mathrm{obs}}$ is shorter than the true radiative lifetime $\tau_{\mathrm{rad}}$ due to collisional quenching of the $\Dfivehalf$ state by residual background-gas molecules~\cite{MadejSankey1990}:
\begin{equation}
  \frac{1}{\tau_{\mathrm{rad}}} = \frac{1}{\tau_{\mathrm{obs}}} - \beta\, P,
  \label{eq:pressure_correction}
\end{equation}
where $\beta$ is the quenching rate coefficient per unit pressure (\si{\per\second\per\milli\bar}) and $P$ is the background-gas pressure.
Determining $\beta$ requires repeating the measurement at several pressures and extrapolating $1/\tau_{\mathrm{obs}}$ to $P = 0$. This was not performed in the present campaign, so \cref{eq:lifetime_weighted} serves only as a lower limit on $\tau_{\mathrm{rad}}$. What can be quantified is the size of the missing quenching rate. Against the most precise of the published values listed in \cref{tab:literature_lifetime}, $\tau = \SI{0.3908(16)}{\second}$ by Letchumanan~\textit{et al.}~\cite{Letchumanan2005}, our observed decay rate is larger by
\begin{equation}
  \frac{1}{\SI{0.325}{\second}} - \frac{1}{\SI{0.3908}{\second}} \simeq \SI{0.52}{\per\second},
  \label{eq:missing_rate}
\end{equation}
some \SI{17}{\percent} of the total observed depopulation rate.

As established in \cref{subsec:storage}, the residual gas at the trap originates mainly in the ion-source chamber. Its composition was not measured, and since the load is set by permeation through an elastomer seal, we take it to be that of air. In classical capture theory~\cite{Langevin1905,GioumousisStevenson1958} the ion--neutral collision rate constant is independent of the collision velocity,
\begin{equation}
  k_{\mathrm{L}} = 2\pi e\sqrt{\frac{\alpha}{4\pi\varepsilon_{0}\,\mu}},
  \label{eq:langevin_rate}
\end{equation}
with $\alpha$ the polarizability volume of the neutral and $\mu$ the ion--neutral reduced mass. Assuming that every collision depopulates the $\Dfivehalf$ state, the quenching coefficient of a species $i$ is $\beta_i = k_{\mathrm{L},i}/k_{\mathrm{B}}T$, and for a mixture with mole fractions $x_i$ the rates add, $\beta = \sum_i x_i\beta_i$. With the polarizabilities of Ref.~\cite{CRCpolarizability}, N$_2$ and O$_2$, which make up \SI{99}{\percent} of air, give $\beta_i = \num{1.6e7}$ and \SI{1.5e7}{\per\second\per\milli\bar} at \SI{300}{\kelvin}, so that $\beta \approx \SI{1.6e7}{\per\second\per\milli\bar}$. This is of the same order as the coefficients $\beta_{\mathrm{impl}}$ in \cref{tab:literature_lifetime}, implied by the missing rate of \cref{eq:missing_rate} and its counterparts for the other published lifetimes at the trap pressure of \SI{2e-8}{\milli\bar} (\cref{subsec:storage}).

\begin{table}[t]
  \centering
  \caption{Lifetime value of the $\Dfivehalf$ state in \SrIon{} from this work, compared with published results. For each published value $\tau$, the last column gives the quenching coefficient $\beta_{\mathrm{impl}} = (1/\tau_{\mathrm{obs}} - 1/\tau)/P$ implied at $P = \SI{2e-8}{\milli\bar}$.}
  \label{tab:literature_lifetime}
  \begin{tabular*}{\columnwidth}{@{\extracolsep{\fill}}lll@{}}
    \hline\hline
    Experimental (\si{\second}) & Reference & $\beta_{\mathrm{impl}}$ (\si{\per\second\per\milli\bar}) \\
    \hline
    $0.325(5)(4)$ & This work & --- \\
    \hline
    $0.3908(16)$  & Letchumanan~\textit{et al.}~\cite{Letchumanan2005} & \num{3e7} \\
    $0.408(22)$   & Bi\'emont~\textit{et al.}~\cite{Biemont2000}       & \num{3e7} \\
    $0.372(25)$   & Madej \& Sankey~\cite{MadejSankey1990OL}           & \num{2e7} \\
    $0.347(11)$   & Barwood~\textit{et al.}~\cite{Barwood1993}         & \num{1e7} \\
    $0.345(33)$   & Gerz~\textit{et al.}~\cite{Gerz1987}               & \num{0.9e7} \\
    \hline\hline
  \end{tabular*}
\end{table}

\section{\label{sec:Discussion}DISCUSSION AND FUTURE PLANS}

The measurements reported in \cref{sec:Meas} show that STRIPE can decelerate, capture and Doppler-cool fast \SI{10}{\kilo\electronvolt} Sr$^+$ beams into well-defined Coulomb crystals. Fluorescence appeared within a few hundred \si{\milli\second} of injection, marking the onset of cooling, while a well-ordered Coulomb crystal formed over several seconds. The detection rate of the photon collection region was measured to be $\eta_\mathrm{PCR} = \SI{296(9)}{photons\,\second^{-1}\,ion^{-1}}$, for a geometric collection efficiency of about \SI{0.25}{\percent} of the full solid angle, and the cold-ion storage lifetime was $\tau_{\mathrm{trap}} = \SI{11.6(10)}{\minute}$, likely limited by background pressure. The dipole-forbidden $\Shalf\to\Dfivehalf$ transition was then used as a probe to characterize the observed linewidth and assess possible broadening mechanisms. To investigate these, the ions were interrogated in two configurations, which differ in both the trapping position and the direction of the probe beam. The narrower line was obtained in the within-electrode configuration with the probe along the trap axis, with a Zeeman-resolved Gaussian linewidth $\FWHMG = \SI{14.5(8)}{\mega\hertz}$, attributed mainly to residual micromotion. The lifetime of the metastable $\Dfivehalf$ state was also measured, giving $\tau_{\mathrm{obs}} = 0.325(5)_{\mathrm{stat}}(4)_{\mathrm{sys}}\,\si{\second}$ without correcting for collisional quenching. These results are summarized in \cref{tab:summary}.

\begin{table*}[t]
  \centering
  \caption{Summary of the commissioning numbers reported in \cref{sec:Meas}.}
  \label{tab:summary}
  \begin{tabular*}{\textwidth}{@{\extracolsep{\fill}}lll@{}}
    \hline\hline
    Quantity & Value & Section \\
    \hline
    PCR detection rate $\eta_\mathrm{PCR}$                       & \SI{296(9)}{photons\,\second^{-1}\,ion^{-1}}  & \cref{subsec:LCR} \\
    Storage lifetime $\tau_{\mathrm{trap}}$            & \SI{11.6(10)}{\minute}                        & \cref{subsec:storage} \\
    Within-electrode Gaussian FWHM $\FWHMG$            & \SI{14.5(8)}{\mega\hertz}                      & \cref{subsubsec:config2} \\
    $\Dfivehalf$ lifetime $\tau_{\mathrm{obs}}$ (uncorrected) & $0.325(5)_{\mathrm{stat}}(4)_{\mathrm{sys}}\,\si{\second}$           & \cref{subsubsec:lifetime_D52} \\
    \hline\hline
  \end{tabular*}
\end{table*}

The demonstrated applicability of STRIPE to beams under conditions similar to those at radioactive ion beam facilities has motivated a new collaboration with the IGISOL facility~\cite{MOORE2013208}, aimed at in-trap LS of radioactive isotopes. An upgraded trap based on STRIPE and the findings of this work, built with the ideal ratio $r_{\mathrm{rod}}/r_{0} = 1.12$ (\cref{sec:Setup}), is being developed there, where trapping and laser cooling of \SI{2}{\kilo\electronvolt} Sr$^+$ beams have recently been demonstrated. A separate publication based on this new setup will follow in the near future. At IGISOL, the new trap will further be loaded from the facility's radio-frequency quadrupole (RFQ) cooler-buncher~\cite{Nieminen2001}. In addition to delivering ions in short bunches with reduced transverse emittance, improving injection and deceleration efficiency into the trap, the cooler-buncher compresses the longitudinal energy spread of the beam to below \SI{1}{\electronvolt}. The narrower velocity distribution entering the trap should lead to a better overlap with the frequency detuning of the \SI{422}{\nano\metre} cooling laser and faster crystallization.

The linewidth currently achieved with STRIPE on a narrow transition, about \SI{15}{\mega\hertz}, is already below or on par with the tens-of-\si{\mega\hertz} linewidths of the dipole transitions used for LS on fast beams~\cite{Cocolios2013, Karadimas2026}. Planned upgrades to suppress the residual axial micromotion --- for example, individual control of the two applied RF phases and/or RF pickup filters --- are expected to bring the micromotion broadening below \SI{1}{\mega\hertz}. Even then, the width of, e.g., the $\Shalf \to \Dfivehalf$ line would be of order \SI{1}{\mega\hertz} at the Doppler limit~\cite{Wineland1979}, an order of magnitude above the radial secular frequency of about \SI{100}{\kilo\hertz}, making resolved-sideband cooling impractical. Suppressing the micromotion broadening below \SI{1}{\mega\hertz} is also challenging in our centimeter-scale trap, since any component of the probe laser in the radial plane makes the line sensitive to radial micromotion as well. The latter is easier to suppress in smaller traps driven at higher RF frequency due to their stiffer radial confinement. We therefore pursue two complementary routes. First, in parallel with STRIPE, a dedicated precision ion-trap setup~\cite{Lassegues2026} has been under development in the group, combining a much smaller inner-electrode half-distance $r_{0}$ with an RF drive frequency of order \SI{10}{\mega\hertz}. Together these place its secular frequencies at the \si{\mega\hertz} level, allowing sideband cooling down to the motional ground state. A copy of this trap is foreseen alongside STRIPE and the IGISOL trap, with cold ions transferred from the latter to the smaller precision one through a linear ion guide, already designed on the basis of SIMION~\cite{Dahl2000} simulations. Second, both STRIPE and the IGISOL trap will be used for double-resonance RF spectroscopy. Motional contributions to a linewidth scale with the transition frequency, so that the \SI{14.5}{\mega\hertz} measured on the \SI{674}{\nano\metre} line corresponds to a few tens of \si{\hertz} on a \si{\giga\hertz}-scale transition. Systematic effects that do not scale in this way, such as magnetic-field contributions, remain to be characterized. Doppler-cooling the ions of interest will ensure that the levels driven remain sufficiently resolved from one another to extract the hyperfine constants, including octupole terms and nuclear $g$-factors, though not isotope shifts.

In parallel with the laser and RF spectroscopy plans, a further collaboration at IGISOL aims to deliver cold ions, cooled either by laser or by sympathetic cooling, from the new trap to the JYFLTRAP double Penning trap~\cite{Eronen2012}, located a few meters downstream. Simulation studies performed by the MIRACLS collaboration at ISOLDE/CERN~\cite{PhysRevResearch.4.033229} have shown that the mass-measurement precision of the phase-imaging ion-cyclotron-resonance (PI-ICR) technique~\cite{Eliseev2013}, in regular use at IGISOL~\cite{Nesterenko2018}, can be improved by up to two orders of magnitude by cooling the ions to the Doppler limit, as long as other uncertainty contributions such as field misalignments and instabilities are negligible.

\begin{acknowledgments}
We thank Paul Van den Bergh for his help in setting up the experiment, and the electronic and mechanical workshops of KU Leuven for delivering the parts needed for the setup and for their on-site assistance. We also thank Franziska Maria Maier for valuable comments on the manuscript. This work was supported by KU Leuven (Grant No. C14/22/104) and Fonds Wetenschappelijk Onderzoek (Project No. G0A0723N and Odysseus Project No. G0F7321N).
\end{acknowledgments}

\section*{Author Declarations}
\subsection*{Conflict of Interest}
The authors have no conflicts to disclose.

\subsection*{Author Contributions}
\textbf{Stefanos Pelonis}: Formal analysis (lead); Investigation (equal); Methodology (equal); Software (lead); Visualization (lead); Writing – original draft (lead); Writing – review \& editing (equal).
\textbf{Tobias Christen}: Writing – review \& editing (equal).
\textbf{Wouter Gins}: Software (supporting); Writing – review \& editing (equal).
\textbf{Bram Hendrickx}: Investigation (supporting); Writing – review \& editing (equal).
\textbf{Phillip Imgram}: Conceptualization (equal); Investigation (equal); Methodology (equal); Project administration (supporting); Software (supporting); Supervision (supporting); Writing – review \& editing (equal).
\textbf{Angelos Karadimas}: Investigation (supporting); Writing – review \& editing (equal).
\textbf{\'Agota Koszor\'us}: Conceptualization (supporting); Funding acquisition (supporting); Methodology (supporting); Project administration (supporting); Writing – review \& editing (equal).
\textbf{Sandro Kramer}: Investigation (supporting); Writing – review \& editing (equal).
\textbf{Stef Kunnen}: Investigation (supporting); Writing – review \& editing (equal).
\textbf{Pierre Lassegues}: Investigation (supporting); Writing – review \& editing (equal).
\textbf{Andrea Raggio}: Investigation (supporting); Writing – review \& editing (equal).
\textbf{Robbe Van Duyse}: Investigation (supporting); Writing – review \& editing (equal).
\textbf{Ruben P. de Groote}: Conceptualization (equal); Funding acquisition (lead); Methodology (equal); Project administration (lead); Software (supporting); Supervision (lead); Writing – review \& editing (equal).

\section*{Data Availability Statement}

The data that support the findings of this study are available from the corresponding author upon reasonable request.

\appendix

\section{\label{app:zeeman}Zeeman-resolved electric-quadrupole spectroscopy model}

This appendix presents the spectral model used in \cref{subsubsec:config2} to fit the
Zeeman-resolved \Shalf\ $\to$ \Dfivehalf\ electric-quadrupole (E2) spectra recorded in
the within-electrode configuration. We first write down the complete model and then
derive each of its ingredients in turn.

For a magnetic-field magnitude $B$, the fluorescence-dip signal is modeled as
\begin{align}
  F(\nu;\,B) &= -\mathcal{A}\sum_{m_{S},\,m_{D}} W_{q}\,S_{\mathrm{at}}(m_{S},m_{D})\, \nonumber\\
  &\quad\times V\!\bigl(\nu;\,\nu_{0}+\Delta\nu(m_{S},m_{D};B),\,\FWHMG,\,\Gamma_{L}\bigr) \nonumber\\
  &\quad + b_{0} + b_{1}\nu .
  \label{eq:app_spectrum}
\end{align}
The sum runs over the ten allowed Zeeman components
$\ket{\Shalf,m_{S}}\to\ket{\Dfivehalf,m_{D}}$ connecting the two levels, as set by the E2 selection rule $\Delta m \equiv m_{D}-m_{S}\in\{0,\pm1,\pm2\}$.
Each enters as a Voigt profile $V$ centered at the field-dependent frequency
$\nu_{0}+\Delta\nu(m_{S},m_{D};B)$, where $\nu_{0}$ is the field-free line center and $\Delta\nu$ the Zeeman shift. Its area is the product of an atomic line strength
$S_{\mathrm{at}}(m_{S},m_{D})$ and a geometric weight $W_{q}$ ($q=m_{D}-m_{S}$) that encodes the orientation of the laser polarization and propagation relative to the field.
The overall scale $\mathcal{A}>0$ and the linear baseline $b_{0}+b_{1}\nu$ are free per dataset, and the leading negative sign reflects the shelving-induced drop in fluorescence on resonance.

The clock transition connects \Shalf\ ($J=1/2$, $g_{S}=2.002256$) to \Dfivehalf\
($J=5/2$, $g_{D}=1.200334$). Since $^{88}$Sr$^+$ has zero nuclear spin, the levels carry no hyperfine structure and are labeled solely by $J$ and $m_{J}$. In a static field $\vb{B}$ each sublevel $\ket{J,m_{J}}$ shifts by $g_{J}\mu_{B}m_{J}B$, so a component is
displaced from $\nu_{0}$ by
\begin{equation}
  \Delta\nu(m_{S},m_{D}) = \frac{\mu_{B}B}{h}\bigl(g_{D}m_{D}-g_{S}m_{S}\bigr).
  \label{eq:app_zeeman_shift}
\end{equation}

The relative strength of each component follows from the E2 transition rate, which is
proportional to $|\bra{J',m'}\hat{T}^{(2)}_{q}\ket{J,m}|^{2}\,W_{q}$ with $q=m'-m$. The
Wigner--Eckart theorem separates the atomic matrix element into a Wigner $3j$ symbol and
a reduced matrix element common to all components,
\begin{equation}
  \bra{J',m'}\hat{T}^{(2)}_{q}\ket{J,m}
    = (-1)^{J'-m'}\begin{pmatrix} J' & 2 & J \\ -m' & q & m \end{pmatrix}
      \bra{J'}\!\|\hat{T}^{(2)}\|\!\ket{J}.
  \label{eq:app_wigner_eckart}
\end{equation}
As only relative strengths are needed, the reduced matrix element cancels and the atomic
factor is the squared $3j$ symbol,
\begin{equation}
  S_{\mathrm{at}}(m_{S},m_{D})
    = \begin{pmatrix} 5/2 & 2 & 1/2 \\ -m_{D} & q & m_{S} \end{pmatrix}^{2},
    \qquad q=m_{D}-m_{S}.
  \label{eq:app_Sat}
\end{equation}

As the transition is E2, it couples to the field gradient $\nabla\vb{E}$ and hence
involves the photon wave vector $\hat{k}$ alongside the polarization $\hat{\varepsilon}$.
The weight of the spherical component $q$ is the squared modulus of the rank-2 tensor built
from $\hat{\varepsilon}$ and $\hat{k}$,
\begin{equation}
  W_{q} = \Bigl|\bigl(\hat{\varepsilon}\otimes\hat{k}\bigr)^{(2)}_{q}\Bigr|^{2}
        = \Bigl|\sum_{m_{1},m_{2}=-1}^{+1}
            \langle 1,m_{1};1,m_{2}|2,q\rangle\,\varepsilon_{m_{1}}k_{m_{2}}\Bigr|^{2},
  \label{eq:app_Wq}
\end{equation}
evaluated in the $B$-frame ($\hat{z}'\parallel\vb{B}$). In our geometry, the laboratory
frame is fixed by $\hat{z}_{\mathrm{lab}}\parallel\hat{k}$ and
$\hat{y}_{\mathrm{lab}}\parallel\vb{B}$, taking $\vb{B}$ along the coil axis, and the polarization analyzer gives an azimuth
$\varphi=\SI{12.8}{\degree}$ from $\hat{x}_{\mathrm{lab}}$, i.e.,
$\hat{\varepsilon}=0.9751\,\hat{x}+0.2215\,\hat{y}$. A central feature of this
configuration, in which $\hat{k}\perp\vb{B}$, is that $W_{0}=0$ exactly, independently of
the polarization state: in the $B$-frame $\hat{k}$ lies along $\hat{x}'$ so $k_{0}=0$,
while $\hat{\varepsilon}\perp\hat{k}$ forces $\varepsilon_{x'}=0$. In \cref{eq:app_Wq} the $m_{1}=m_{2}=0$
term vanishes with $k_{0}$, and the two remaining $q=0$ terms ($m_{1}=-m_{2}=\pm1$) sum to
$-\varepsilon_{x'}k_{x'}/\sqrt{6}$, which vanishes with $\varepsilon_{x'}$. All $\Delta m=0$
components are therefore geometrically suppressed. For $\varphi=\SI{12.8}{\degree}$ the
surviving weights are $W_{\pm2}=2.377\times10^{-1}$ (\SI{95.1}{\percent} of
$\sum_{q}W_{q}$) and $W_{\pm1}=1.227\times10^{-2}$ (\SI{4.9}{\percent}). A stray-field component off the coil axis tilts the quantization axis and lifts the suppression of the $\Delta m=0$ components; its effect on the fit is quantified in \cref{subsubsec:config2}.

Finally, each component is a Voigt profile of shared Gaussian width \FWHMG\ and Lorentzian
width $\Gamma_{L}$, evaluated through the Faddeeva function $w(z)$ as
$V(\nu;\nu_{0},\FWHMG,\Gamma_{L})=\mathrm{Re}[w(z)]/(\sigma\sqrt{2\pi})$ with
$z=[(\nu-\nu_{0})+i\gamma]/(\sigma\sqrt{2})$, $\sigma=\FWHMG/(2\sqrt{2\ln2})$ and
$\gamma=\Gamma_{L}/2$. The three spectra at $I \in \{1, 2, 3\}\,\si{\ampere}$ are fitted
simultaneously with \cref{eq:app_spectrum}: the Gaussian width \FWHMG\ is shared across
datasets, the Lorentzian width is fixed to $\Gamma_{L}=0$ and the azimuth to
$\varphi=\SI{12.8}{\degree}$, while the field magnitude $B_{i}$, the scale
$\mathcal{A}_{i}$ and the baseline coefficients $b_{0,i},b_{1,i}$ are free for each
dataset. The minimization uses Levenberg--Marquardt least squares as implemented in
\texttt{lmfit}.

\section{\label{app:lcr_model}Detection-rate model}

This appendix gives the simplified model used in \cref{subsec:LCR} to estimate the photon rate $R_\mathrm{sc}$ scattered per ion, which enters \cref{eq:Rth}. The \SrIon{} cooling cycle is treated as a three-level system comprising the $\Shalf$, $\Phalf$ and $\Dthreehalf$ states, driven by the \SI{422}{\nano\metre} cooling and \SI{1092}{\nano\metre} repump lasers. We assume a Gaussian transverse intensity profile, for which the on-axis intensity of a beam of power $P$ and $1/e^{2}$ diameter $d$ is
\begin{equation}
  I_{0} = \frac{8P}{\pi d^{2}},
  \qquad
  s_{0} = \frac{I_{0}}{I_\mathrm{sat}},
  \label{eq:intensity}
\end{equation}
with $I_\mathrm{sat}=\pi h c\,\Gamma/3\lambda^{3}\approx \SI{37.5}{\milli\watt\per\centi\metre\squared}$ at \SI{422}{\nano\metre}, where $\Gamma/2\pi=\SI{21.5}{\mega\hertz}$ is the natural linewidth of the cooling transition, derived from the measured $\Phalf$ lifetime of \SI{7.39(7)}{\nano\second}~\cite{Pinnington1995}. Since the Doppler width at the temperature estimate of \cref{subsec:cooling} is small compared with both $\Gamma$ and the cooling detunings, a Lorentzian lineshape is adopted. The stimulated rate $R_\mathrm{st}$ on each transition ($R_{422}$ or $R_{1092}$), summed over the two \SI{422}{\nano\metre} components $j$ at detunings $\delta_{j}$, is
\begin{equation}
  R_\mathrm{st} = \frac{\Gamma}{2}\sum_{j}
      \frac{s_{0,j}}{1+\left(2\delta_{j}/\Gamma\right)^{2}},
\end{equation}
and the steady-state population of the $\Phalf$ state follows from the three-level rate equations as
\begin{equation}
  n_{P} = \left( 3 + \frac{\Gamma_{SP}}{R_{422}}
                   + \frac{\Gamma_{DP}}{R_{1092}} \right)^{-1},
  \label{eq:nP}
\end{equation}
where $\Gamma_{SP}$ and $\Gamma_{DP}$ are the partial decay rates of the $\Phalf$ state to the $\Shalf$ and $\Dthreehalf$ levels, with $\Gamma_{SP}+\Gamma_{DP}=\Gamma$ and branching ratio $\Gamma_{SP}:\Gamma_{DP}\approx94.5:5.5$~\cite{Likforman2016}. The total \SI{422}{\nano\metre} photon rate emitted into $4\pi$ is then $R_\mathrm{sc}=\Gamma_{SP}\,n_{P}$.

\bibliography{aipsamp}

\end{document}